\documentclass[12pt]{iopart}
\usepackage{iopams}
\expandafter\let\csname equation*\endcsname\relax

\expandafter\let\csname endequation*\endcsname\relax

\usepackage{amsmath}
\usepackage{amstext,amsthm}
\usepackage{esint}
\usepackage{graphicx}
\usepackage[colorlinks,linkcolor=blue,urlcolor=blue,citecolor=blue]{hyperref}
\usepackage{bm,bbm}
\usepackage[normalem]{ulem}

\begin{document}

\newcommand{\be}{\begin{equation}}
\newcommand{\ee}{\end{equation}}
\newcommand{\ben}{\begin{equation}}
\newcommand{\een}{\end{equation}}
\newcommand{\barr}{\begin{eqnarray}}
\newcommand{\earr}{\end{eqnarray}}
\def\lf{\left\lfloor}   
\def\rf{\right\rfloor}

\theoremstyle{plain}
\newtheorem*{thm}{Theorem}

\theoremstyle{definition}
\newtheorem*{notation}{Notation}

\newcommand{\R}{\mathbb{R}}
\newcommand{\de}{\mathrm{d}}
\renewcommand{\rho}{\varrho}
\newcommand{\avg}[1]{\left< #1 \right>} 
\newcommand{\ddet}[1]{\underset{#1}{\operatorname{det}}}
\newcommand{\pper}[1]{\underset{#1}{\operatorname{per}}}
\newcommand{\detalpha}[2]{\underset{{#1}}{\operatorname{det}_{#2}}}
\newcommand{\inter}[2]{\left[ #1\,.\, .\, #2\right)}
\newcommand{\Ai}{\operatorname{Ai}}

\title{Free fermions and $\alpha$-determinantal processes}

\author{Fabio Deelan Cunden$^{1}$, Satya N. Majumdar$^{2}$, Neil O'Connell$^{1}$}
\address{$^{1}$ School of Mathematics and Statistics, University College Dublin, Belfield, Dublin 4, Ireland\\
$^{2}$ LPTMS, CNRS, Univ. Paris-Sud, Universit\'e Paris-Saclay, 91405 Orsay, France }

\date{\today}

\begin{abstract} 
The $\alpha$-determinant is a one-parameter generalisation of the standard determinant, 
with $\alpha=-1$ corresponding to
the determinant, and $\alpha=1$ corresponding to the permanent. In this paper
a simple limit procedure to construct $\alpha$-determinantal point processes out of fermionic 
processes is examined.  The procedure is illustrated for a model of $N$
free fermions in a harmonic potential. When the system is in the ground state, 
the rescaled correlation functions converge for large $N$ to determinants (of the sine
kernel in the bulk and the Airy kernel at the edges). 
We analyse the point processes associated to a special family of excited states
of fermions and show that appropriate scaling limits generate $\alpha$-determinantal processes. 
Links with wave optics and other random matrix models are suggested.
\end{abstract}

\maketitle

\section{Introduction}
\par
Determinantal and permanental processes are point processes whose correlation functions $\rho_{n}(x_1,\dots,x_n)$ exist for all $n\in\mathbb{N}$, and are given by 
\be
\rho_{n}(x_1,\dots,x_n)=
\begin{cases}
\ddet{1\leq i,j\leq n}K(x_i,x_j)\quad&\text{(determinantal)}\\
\pper{1\leq i,j\leq n}K(x_i,x_j)\quad&\text{(permanental)}.
\end{cases}
\ee
The function $K(x,y)$ is called \emph{correlation kernel} and can be thought of as the integral kernel of some integral operator. There is no need for us to review the history and ubiquity of determinantal and permanental processes in mathematical physics and probability~\cite{Macchi75,Mehtabook,Soshnikov00,Torquato08}. 
Another, perhaps not so well-known class of processes are the so-called $\alpha$-determinantal processes. The $\alpha$-determinant  of a $n\times n$ matrix $A$ is
\be
\detalpha{}{\alpha}A=\sum_{\sigma\in S_n}\alpha^{n-m(\sigma)}A_{\sigma(1)1}A_{\sigma(2)2}\cdots A_{\sigma(n)n}
\ee
where $m(\sigma)$ is the number of disjoint cycles in the permutation $\sigma$ --- thus, for example, the identity permutation, corresponding to the term $A_{11}A_{22}\cdots A_{nn}$ contains $n$ cycles and appears with weight $\alpha^0$, whereas the term $A_{12}A_{23}\cdots A_{n1}$, corresponding to a single cycle appears with weight $\alpha^{n-1}$. Namely, we simply replace the signature $\operatorname{sgn}(\sigma)=(-1)^{n-m(\sigma)}$ by $\alpha^{n-m(\sigma)}$ in the definition of the ordinary determinant $\det A$.

It is clear that 
\be
\operatorname{det}_{-1}A=\det A,\quad \operatorname{det}_{1}A=\operatorname{per} A,\quad \operatorname{det}_0A=A_{11}A_{22}\cdots A_{nn}.
\ee
\par
Vere-Jones~\cite{Vere-Jones88,Vere-Jones97} introduced $\alpha$-determinants to treat the probability density functions of multivariate binomial and negative binomial distributions in a unified way.
Later, Shirai and Takahashi~\cite{Shirai03} utilised the $\alpha$-determinant to define a parametric family of point processes which extend the fermionic and bosonic point processes. Let $\alpha\in\mathbb{R}$ and $K$ a kernel from say $\mathbb{R}^2$ to $\mathbb{C}$.  An $\alpha$-determinantal point process with kernel $K$  is defined, when it exists, as the point process with $n$-point correlation functions ($n\geq 1$)
\be
\rho_n(x_1,\dots,x_n)=\detalpha{1\leq i,j\leq n}{\alpha}K(x_i,x_j).
\ee
The values $\alpha=-1$ and $\alpha=1$ correspond to determinantal and permanental processes, respectively. The case $\alpha=0$ corresponds to the Poisson process with intensity $K(x,x)$.
\par
Several authors have established necessary and sufficient conditions for the existence of $\alpha$-determinantal processes. See~\cite{Maunoury15} and references therein. 
\par
In this paper, we shall only be concerned with the case $\alpha<0$; in this case, a necessary condition for existence is that  is that $-\frac{1}{\alpha}\in\mathbb{N}$
 (otherwise the $\alpha$-determinants $\det_{\alpha}K(x_i,x_j)$ can be negative). 
If  $-\frac{1}{\alpha}\in\mathbb{N}$,  and $K$ is self-adjoint with $0\leq K\leq -\frac{1}{\alpha}$, then the $\alpha$-determinantal process exists. In fact, it is just a union (or `superposition') of $-\frac{1}{\alpha}$ i.i.d. copies 
of the determinantal process with kernel $-\alpha K$. 
\par
Although $\alpha$-determinantal processes have been investigated theoretically, concrete realisations of them have not been discussed as much in the literature.
The present paper might be thought of as a first step in this direction; hopefully more examples will emerge in time. 
\par
The purpose of this paper is
to provide an explicit construction of $\alpha$-determinantal point processes
as limiting cases arising naturally in a model of $N$ non-interacting
fermions in a one-dimensional harmonic potential. 
We consider a family of many-body excited states parametrized by
a real number $a$, where $a=0$ corresponds to the fermionic ground state. The associated determinantal process is a \emph{block projection process}. The first observation of the paper is
that, as the parameter varies from $a=0$ to $a\to \infty$, the average density of fermions  crosses over
from the Wigner semicircular distribution (in the quantum ground state) to the arcsine distribution (corresponding to a fully `classical' excited state); this is
consistent with the correspondence principle of quantum mechanics.
The main result of the paper is that if the limit $a\to \infty$ is taken
appropriately, then the block projection process associated to the many-body excited state converges weakly (in the scaling limit)
to an $\alpha$-determinantal process with $\alpha=-1/2$. In the same setting, we also
provide the explicit construction
of $\alpha$-determinantal processes for general $\alpha=-1/m$, with $m\in\mathbb{N}$.
These results are summarised as a Theorem in Section~\ref{sec:sum}.
\par
The outline of the paper is as follows. In the next section we record the spectral properties of non-interacting fermions in a harmonic potential. In Section~\ref{sec:det}, we first recall the connection between free fermions in the ground state and the GUE processes, and some immediate implications of this connection; then, we introduce a first example of block projection process and we analyse its scaling limits and the convergence to an $\alpha$-determinantal process. In Section~\ref{sec:block} we generalise the construction of block projection processes and show their convergence to $\alpha$-determinantal processes (superposition of sine processes). A summary of the main result - weak convergence of block projection fermionic processes to $\alpha$-determinantal processes, further remarks and links with wave optics and random matrices conclude the paper (Section~\ref{sec:sum}).
\par
\paragraph{Some notation.} For $a< b$ we use the notation $\inter{a}{b}$ to denote the integer interval $\{\lf{a}\rf,\lf{a}\rf+1,\ldots,\lf{b}\rf-1\}$. For $x_i,x_j\in\R$ we write $x_{ij}=x_i-x_j$. Denote the complex conjugate of $z$ by $\overline{z}$.

\par
\section{Free fermions in a harmonic potential and determinantal processes}
\label{sec:fermions}

The connection between free fermions and determinantal processes has been known for a long time~\cite{Forrester_book,Macchi75,Macchi77,Mehtabook,Soshnikov00}. This
connection has been used in various contexts, such as in the analysis of a class of matrix models (Moshe-Neuberger-Shapiro 
model)~\cite{Johansson07,MNS94}, in the study of non-intersecting step-edges on a crystal~\cite{Einstein03}, and
in establishing a connection between non-intersecting Brownian interfaces in the presence of a confining potential and Wishart
random matrices~\cite{NM09}. However, in the specific context of $N$ non-interacting fermions trapped in a one-dimensional 
harmonic potential, the connection to the Gaussian unitary ensemble (GUE) was established and used only recently in a series of
papers: first somewhat indirectly in Ref.~\cite{Eisler13,Vicari12}, and then more explicitly in Ref.~\cite{Eisler13b,MMSV14} in the context of
full counting statistics of fermions. Later, this connection has been further exploited quite heavily in calculating various physical properties of $1$-d trapped fermions, such as the correlation functions near
the edges of the trapped Fermi gas~\cite{DLMS15,DLMS16,LMS18}, effects of finite temperature and 
the connection to the Kardar-Parisi-Zhang equation at finite time~\cite{DLMS15,DLMS16},   
computation of the number variance, other linear statistics~\cite{GMST18,GMS17,MMSV14,MMSV16} and the 
entanglement entropy~\cite{CLM15}. Free fermions in a one-dimensional 
non-harmonic traps, singular or with hard edges such as a box potential 
(where the determinantal process is not GUE), 
have also been studied~\cite{LLMS17,LLMS18}. 
In particular, the relationship between fermions in a box with different boundary conditions
and the classical compact groups have been explored~\cite{CMO18,FMS11}.
For a review of some of these recent developements in the physics literature, 
see Ref.~\cite{DLMSreview}. 
In this section, we first recall the precise connection between
the {\it ground state} of non-interacting fermions in a harmonic potential and the 
GUE determinantal process 
and then extend this to a class of {\it special excited} states that, in a certain appropriate limit of high energy, converges to
 $\alpha$-determinantal process with $\alpha=-1/2$.

Denote by $\psi_k(x)$ the Hermite wavefunctions
\be
\psi_k(x)=h_k(x)e^{-x^2/4},\qquad h_k(x)=\frac{(-1)^k}{\sqrt{\sqrt{2 \pi } k!}}\; e^{x^2/2} \frac{\partial ^k}{\partial x^k}e^{-x^2/2},\qquad k=0,1,2,\dots.
\ee
They are solutions of the Schr\"odinger equation 
\be
\left(-\frac{\partial^2}{\partial x^2}+\frac{x^2}{4}\right)\psi_{k}(x)=\left(k+\frac{1}{2}\right)\psi_{k}(x),\quad x\in\R,
\ee
and form a complete orthonormal system in $L^2(\mathbb{R};\de x)$. 
Physically, $\psi_k(x)$ is an eigenfunction of the quantum harmonic oscillator corresponding to the eigenvalue $E_k=k+1/2$, i.e. the eigenstate of a quantum particle in a harmonic potential at the energy level $E_k$. 
\par
The normalised eigenstates $\Psi_{k_1,\dots,k_N}(x_1,\dots,x_N)$ of a system of $N$ spin-polarized fermions in the same harmonic potential, are given by  antisymmetric linear combinations of the $\psi_k$'s, and can be conveniently written as Slater determinants
\be
\Psi_{k_1,\dots,k_N}(x_1,\dots,x_N)=\frac{1}{\sqrt{N!}}\det_{1\leq i,j\leq N} \psi_{k_i}(x_j),\qquad \text{with $0\leq k_1<k_2<\cdots<k_N$}.
\ee
They are eigenfunctions of the operator $\sum_i\left(-\frac{\partial^2}{\partial x_i^2}+\frac{x_i^2}{4}\right)$ with eigenvalues $E=k_1+\cdots+k_N+N/2$, in the subspace of completely antisymmetric states $\Psi(x_{\sigma(1)},\dots,x_{\sigma(N)})=\operatorname{sgn}(\sigma)\Psi_{k_1,\dots,k_N}(x_1,\dots,x_N)$. These facts follow from the basic properties of  the determinant. 
\par
The modulus square of the wave function $\Psi(x_1,\dots,x_N)$ can be interpreted as the joint probability density of the particles positions. If we denote $J=\{k_1,\dots,k_N\}\subset\mathbb{N}$, we can write
\be
|\Psi_J(x_1,\dots,x_N)|^2=\frac{1}{N!}\det_{1\leq i,j\leq N}K_J(x_i,x_j),
\ee
where 
\be
K_J(x,y)=\sum_{k\in J}\overline{\psi}_k(x)\psi_k(y)
\ee
is the integral kernel of the projection operator onto the $N$-dimensional subspace  $\operatorname{span}\left\{\psi_k(x)\colon k\in J\right\}\subset L^2(\R;\de x)$. In fact, $|\Psi_J(x_1,\dots,x_N)|^2$ defines a determinantal point process of $N$ particles on $\mathbb{R}$ with respect to $\de x$ with kernel $K_J(x,y)$. The $n$-th correlation function of the process is
\be
\rho_n(x_1,\dots,x_n)=\frac{N!}{(N-n)!}\int|\Psi_J(x_1,\dots,x_N)|^2\de x_{n+1}\cdots\de x_N=\det_{1\leq i,j\leq n}K_J(x_i,x_j).
\ee
\par
The main observation of this paper is that, for special choices of the energy levels $J\subset\mathbb{N}$, the scaling limit in the bulk of the determinantal process defined by $K_J(x,y)$ is an $\alpha$-determinantal process.

\section{Free fermions}
\label{sec:det}
\subsection{Ground state and the GUE eigenvalue process}
Suppose that $J=\inter{0}{N}$ corresponding to the wavefunction
\ben
\Psi_J(x_1,\dots,x_N)=\frac{1}{\sqrt{N!}}\det_{1\leq i,j\leq N} \psi_{i-1}(x_j).
\een
This is the unique ground state of $N$ non-interacting fermions in a harmonic potential  (exactly one fermion in each energy state $k_i=i-1$, $i=1,\dots,N$). The ground state energy is 
\be
E_0=\sum_{k\in\inter{0}{N}}(k+1/2)=\frac{N^2}{2}.
\ee
The kernel
\be
K_J(x,y)=\sum_{k\in\inter{0}{N}}\overline{\psi}_k(x)\psi_k(y)
\label{eq:kernel_GUE}
\ee
coincides with the kernel of the GUE ensemble of random matrix theory. The  determinantal point process on $\mathbb{R}$ defined by  $K_J(x,y)$ above is know as \emph{GUE process}. 
\par
At first, a large $N$ asymptotics of~\eqref{eq:kernel_GUE} seems hopeless since the number of terms in the sum is $N$. However, for the special choice $J=\inter{0}{N}$, one can apply the Christoffel-Darboux formula and rewrite the kernel in the form
\be
K_J(x,y)=\sqrt{N}\;\frac{\psi_N(x)\psi_{N-1}(y)-\psi_{N-1}(x)\psi_{N}(y)}{x-y},
\ee
which is amenable of a large $N$ analysis by means of the Plancherel-Rotach asymptotic expansions of Hermite polynomials.
\par
It is well-known, for instance, that the number density of particles (one-point function) is asymptotic to the 
semicircular law at leading order in $N$
\be
\rho_1(x)=K_J(x,x)\stackrel{N\to\infty}{\sim} \frac{1}{2\pi}\sqrt{(4 N-x^2)_+}, 
\label{eq:semicircle}
\ee
with the normalization $\int \rho_1(x) dx=N$. Moreover, in the scaling limit in the bulk, 
the GUE process converges to the sine process, a determinantal process on $\R$ with translation invariant kernel
\be
\lim_{N\to\infty}\frac{1}{\rho_1(0)}K_J\left(\frac{x}{\rho_1(0)},\frac{y}{\rho_1(0)}\right)=\frac{\sin\pi(x-y)}{\pi(x-y)}=\frac{\sin\pi x\cos\pi y-\cos\pi x\sin\pi y}{\pi(x-y)}.
\label{eq:sinekernel}
\ee
The behaviour of the process at the endpoints $\pm\sqrt{4N}$ of the density is different. At the edges, on the scale $\Or(N^{-1/6})$ of the typical distance between points, the process converges to the Airy process with kernel
\be
\lim_{N\to\infty}\frac{1}{N^{\frac{1}{6}}}K_J\left(\sqrt{4N}+\frac{x}{N^{\frac{1}{6}}},\sqrt{4N}+\frac{y}{N^{\frac{1}{6}}}\right)=\frac{\Ai(x)\Ai'(y)-\Ai'(x)\Ai(y)}{x-y}.
\label{eq:Airykernel}
\ee
\subsection{Excited states, the correspondence principle and $\alpha$-determinants}
\label{sec:excited}
\begin{figure}
\centering
\includegraphics[width=.65\columnwidth]{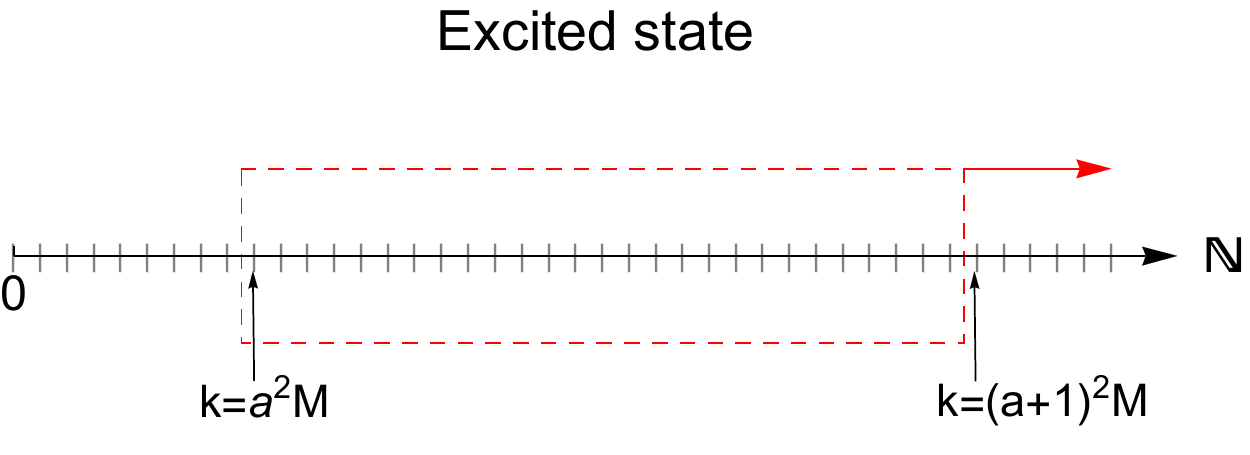}
\caption{ An excited state $J=\inter{a^2M}{(a+1)^2M}$ consisting of $N= ((a+1)^2-a^2)\,M=(2a+1)\,M$ 
consecutive energy levels $(k+1/2)$ of the harmonic oscillator,
starting with $k=a^2\, M$ and ending with $k=(a+1)^2 M-1$. The rectangle denoting the block $J$ is parametrised by $a$, with $a=0$ corresponding to the ground state of $N$ fermions (when $J=\inter{0}{N}$). By increasing $a$, one can slide the block and 
consider a family
of such $J$'s labelled by the single parameter $a$.}
\label{fig:sliding}
\end{figure}

Consider now the case $J=\inter{a^2M}{(a+1)^2M}$, labelling an excited state where $N$ fermions
occupy $N=|J|$ consecutive levels\footnote{We omit, for notational simplicity, to indicate explicitly the integer parts $\lf a^2 M\rf,\dots \lf (a+1)^2 M-1\rf$; we will often do this below without repeating this warning.} $k=a^2M,\ldots, (a+1)^2M-1$  with
$N= ((a+1)^2-a^2)\,M= (2a+1)\,M $. Thus this excited state (or `block' as shown by the rectangle in Fig. \ref{fig:sliding})
is parametrised by $a$, with $a=0$ corresponding to the ground state. The fermions
in this block $J$ forms a determinantal process with kernel
\ben
K_J(x,y)=\sum_{k=a^2M}^{(a+1)^2M-1}\overline{\psi_k}(x)\psi_k(y).
\label{eq:kernel_2block}
\een
Note that this particular way of parametrising the block $J$ (with the starting level $k=a^2\, M= a^2 N/(2a+1)$) turns out
to be useful to express the scaled kernel in a nice and simple way, as is shown later.
\par

\par

\begin{figure}
\centering
\includegraphics[width=.40\columnwidth]{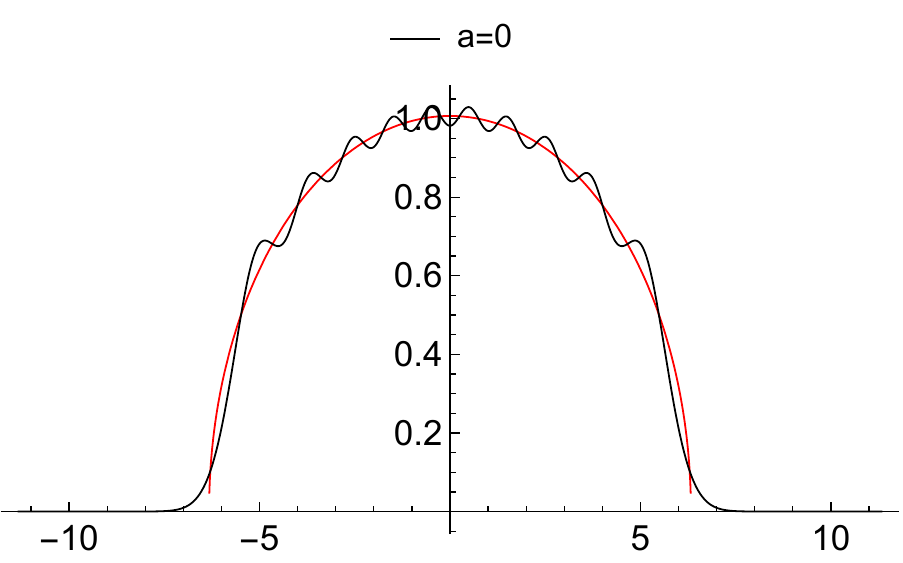}\qquad
\includegraphics[width=.40\columnwidth]{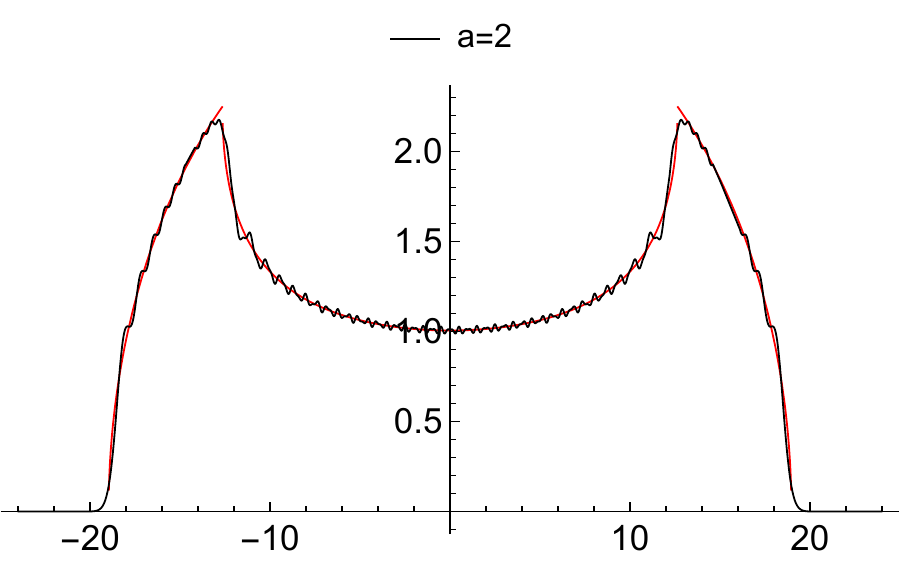}\\
\includegraphics[width=.40\columnwidth]{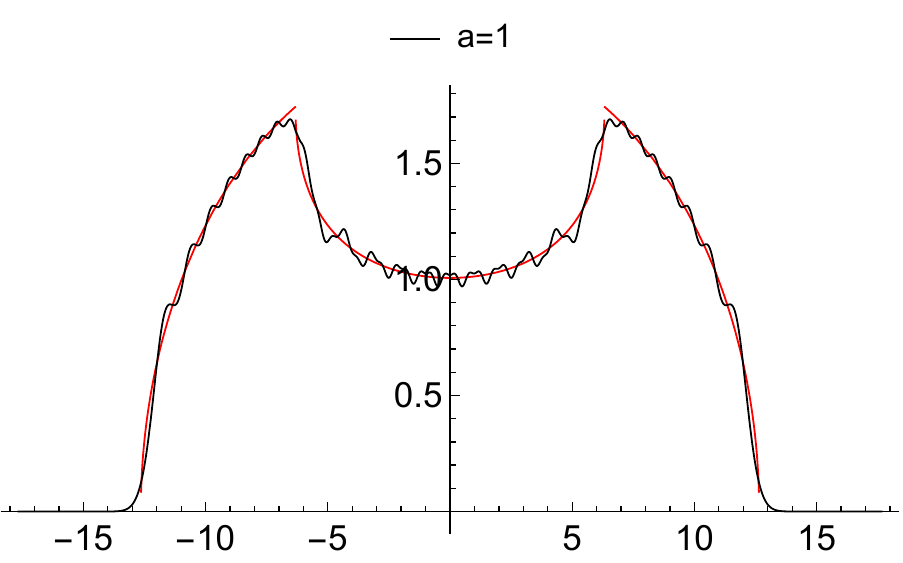}\qquad
\includegraphics[width=.40\columnwidth]{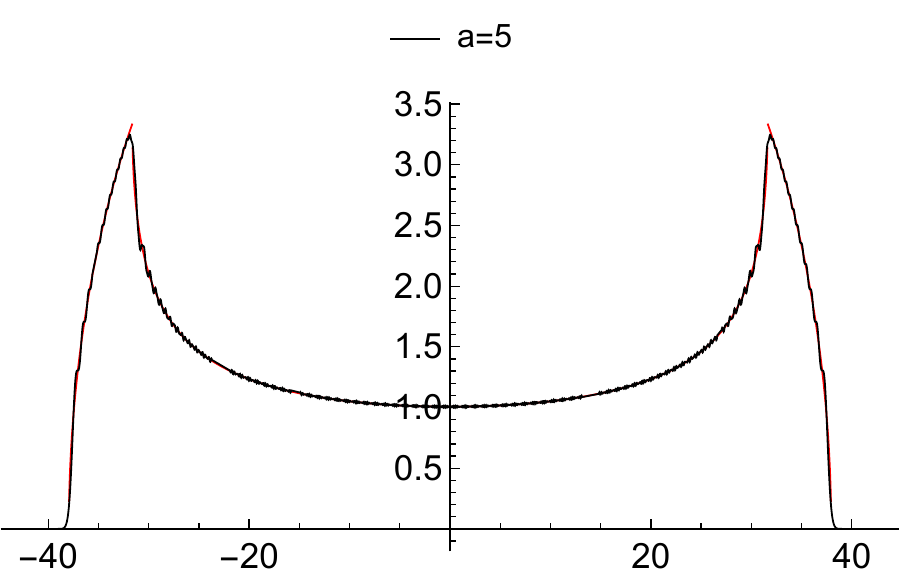}
\caption{Comparison of the one-point functions and the $M\to\infty$ asymptotics~\eqref{eq:one-point_asym}. As $a$ increases the density of states approaches the arcsine law~\eqref{eq:arcsine}.}
\label{fig:one-point}
\end{figure}

We remark that $K_J$ can be written as a (signed) sum of two blocks: 
\be
K_J(x,y)=\sum_{k\in\inter{0}{(a+1)^2M}}\overline{\psi_k}(x)\psi_k(y)-\sum_{k\in\inter{0}{a^2M}}\overline{\psi_k}(x)\psi_k(y).
\label{eq:key}
\ee
This simple observation allows to apply the Christoffel-Darboux formula to both blocks separately, and will be crucial for the following asymptotic analysis.
\paragraph{One-point function}
From~\eqref{eq:key} we can understand easily that the large-$M$ asymptotics of the one-point function (normalised to the number of particles $N=\left(2a+1\right)M$) is
\be
\rho_1(x)=K_J(x,x)\stackrel{M\to\infty}{\sim} \frac{1}{2\pi}\left(\sqrt{(4(a+1)^2 M-x^2)_+}-\sqrt{(4a^2M-x^2)_+}\right).
\label{eq:one-point_asym}
\ee
For $a=0$, so that $N=M$, this reduces to the Wigner semicircular law between $-2\sqrt{M}$ and $2\sqrt{M}$. In general, the one-point function is concentrated between the edges $\pm 2(a+1)\sqrt{M}$. Note that $\lim_{M\to\infty}\rho_1(0)=1/\pi$. The density for a few values of $a>0$ is plotted in Fig.~\ref{fig:one-point}.

For large $a$ the one-point function approaches the arcsine law 
\be
\rho_1(x)\stackrel{a\to\infty}{\sim} \frac{1}{\pi}\frac{(2a+1)M}{\sqrt{4a^2 M-x^2}}1_{|x|<2a\sqrt{M}},
\label{eq:arcsine}
\ee
which is normalized to $\int \rho_1(x)\, dx=(2a+1)M=N$ over
its support $x\in \left[-2a\sqrt{M},\, 2a\sqrt{M}\right]$. The name `arcsine' comes from the fact
that the cumulative number density has the form
\ben
\int_{-\infty}^x \rho_1(x')\,dx'\sim \frac{(2a+1)M}{\pi}\left[\frac{\pi}{2}+ 
\arcsin\left(\frac{x}{2a\sqrt{M}}\right)\right]\, .
\een

A semiclassical explanation for this arcsine law is as follows. The quantum state of a particle can be represented in  the phase space  by a  quasi-probability density known as \emph{Wigner function} (see~\cite{Folland88}). The Wigner function $W_J(x,p)$ associated with the many-body state $\Psi_J(x_1,x_2,\ldots,x_N)$ is
\be
W_J(x,p)= \frac{N}{2\pi} \int_{\R^N}\Psi_J^{*}\left(x+\frac{y}{2}, x_2,\ldots, x_N\right)\Psi_J\left(x+\frac{y}{2}, x_2,\ldots, x_N\right) e^{i p y}\de y \de x_2 \de x_3\ldots \de x_N 
\ee
For large $N$, the Wigner function $W_J(x,p)$ in the phase space is constant in the classically allowed region and zero in the classically forbidden region. The classically allowed region of the phase space is the set of momenta $p$ and positions $x$ such that the energy $E(x,p)=p^2+x^2/4$ is between the lowest occupied energy level $E_{min}=a^2M+1/2$ and the largest occupied level $E_{max}=(a+1)^2M-1/2$. 
\par
Neglecting $o(M)$ terms the region $E_{min}\leq E(x,p)\leq E_{max}$ is the annulus
\be
a^2M\leq p^2+\frac{x^2}{4}\leq (a+1)^2M.
\ee
Therefore, for  large $|J|$, the Wigner function (normalised to the total number of particles) is proportional to the indicator function~\cite{Balazs73,Bettelheim11,Dean18}
\begin{align}
W_J(x,p)dxdp
&\stackrel{M\to\infty}{\sim}\frac{1}{\pi} \mathbbm{1}_{a^2M<p^2+\frac{x^2}{4}<(a+1)^2M}dxdp.
\label{eq:Wigner_f}
\end{align}
The projection of the Wigner function on the $x$-axis gives the average number density:
 $\rho_1(x)=\int W_J(x,p)dp$. When $a=0$, $W_J(x,p)$ is uniform in the ellipse $p^2+\frac{x^2}{4}\leq M$  (a disk if we rescale the axes). The projection of the uniform distribution on the \emph{disk}  is the semicircular law. For $a>1$, $W_J(x,p)$ is uniform in the annulus of radii $a\sqrt{M}$ and $(a+1)\sqrt{M}$, thus explaining the plots in Fig.~\ref{fig:one-point}. For large $a$, the Wigner function $W_J(x,p)$ concentrates on the circle of radius $a\sqrt{M}$, and the projection of the uniform distribution on the \emph{circle} is the arcsine law. We will elaborate more on this key remark in the last section of the paper.
\par
\paragraph{Scaling limits}
The scaling limit in the bulk is
\begin{align}
&\lim_{M\to\infty}\frac{1}{\rho_1(0)}K_{J}\left(\frac{x}{\rho_1(0)},\frac{y}{\rho_1(0)}\right)
=k(x-y),
\end{align}
with
\be
k(x-y)=\frac{1}{\pi (x-y)}\sin\left(\pi(a+1)(x-y)\right)-\frac{1}{\pi (x-y)}\sin\left(\pi a(x-y)\right).
\ee
Using the trigonometric identity $\sin x-\sin y=2\sin\frac{x-y}{2}\cos\frac{x+y}{2}$, the above formula can be rearranged as
\be
k(x-y)=\frac{\sin\frac{\pi}{2} (x-y)}{\frac{\pi}{2} (x-y)}\cos\omega  (x-y).
\label{eq:kernel_ka}
\ee
where we set $\omega=\pi\left(a+1/2\right)$.
For $a=0$ this is, of course,  the sine kernel. 

We will now show that, as $a\to\infty$, the process becomes $\alpha$-determinantal with correlation kernel $\sin\frac{\pi}{2}(x-y)/(\frac{\pi}{2}(x-y))$ and $\alpha=-1/2$. 
\par
First, we remark that for large $a$, the frequency $\omega$ of the cosine factor increases and $k(x-y)$ becomes rapidly oscillating. To get some insight, it is useful to write down explicitly the correlation functions
\ben
\tilde{\rho}_n(x_1,\ldots,x_n)=\lim_{M\to\infty}\frac{1}{\rho_1(0)^{n}}{\rho}_n\left(\frac{x_1}{\rho_1(0)},\ldots,\frac{x_n}{\rho_1(0)}\right)=\det_{1\leq i,j\leq n}k(x_i-x_j).
\een
for the first values of $n$.  
For all $a\geq0$ the one-point function is, of course, constant
\be
\tilde{\rho}_1(x)=1.
\ee
\par
The two-point correlation function is
\ben
\tilde{\rho}_2(x_1,x_2)=1-\left(\frac{\sin\frac{\pi}{2} x_{12}}{\frac{\pi}{2} x_{12}}\right)^2 \cos^2\omega x_{12}
\een
For large $\omega$, the factor $\cos^2\omega x_{12}=\left(1+\cos\omega x_{12}\right)/2$ rapidly oscillates around the mean value $1/2$, so \ben
\lim_{\omega\to\infty}\tilde{\rho}_2(x_1,x_2)=1-\frac{1}{2}\left(\frac{\sin\frac{\pi}{2} x_{12}}{\frac{\pi}{2} x_{12}}\right)^2,
\label{two_point.1}
\een
where this limit is to be understood in the weak sense of integration over compact sets.
In the following, all limits of correlation functions are to be understood in this sense.
\par
The correlation function for three particles is
\begin{align}
\tilde{\rho}_3(x_1,x_2,x_3)&=
1+2\frac{\sin\frac{\pi}{2} x_{12}}{\frac{\pi}{2} x_{12}}\frac{\sin\frac{\pi}{2} x_{23}}{\frac{\pi}{2} x_{23}}\frac{\sin\frac{\pi}{2} x_{31}}{\frac{\pi}{2} x_{31}}
\cos\omega x_{12}\cos\omega x_{23}\cos\omega x_{31}\nonumber\\
&-\left(\frac{\sin\frac{\pi}{2} x_{12}}{\frac{\pi}{2} x_{12}}\right)^2\cos^2\omega x_{12}
-\left(\frac{\sin\frac{\pi}{2} x_{23}}{\frac{\pi}{2} x_{23}}\right)^2\cos^2\omega x_{23}
-\left(\frac{\sin\frac{\pi}{2} x_{31}}{\frac{\pi}{2} x_{31}}\right)^2\cos^2\omega x_{31}.
\end{align}
Again, the squared cosines oscillate around their mean value $1/2$. The product of three cosines can be expanded as
\begin{align}
&\cos\omega x_{12}\cos\omega x_{23}\cos\omega x_{31}\nonumber\\
&=\cos^2\omega x_1\cos^2\omega x_2\cos^2\omega x_3+\sin^2\omega x_1\sin^2\omega x_2\sin^2\omega x_3+\text{zero mean terms},
\end{align}
and thus oscillates around the value $1/8+1/8=1/4$. Therefore 
\begin{align}
\lim_{\omega\to\infty}\tilde{\rho}_3(x_1,x_2,x_3)&=1-\frac{1}{2}\left(\frac{\sin\frac{\pi}{2} x_{12}}{\frac{\pi}{2} x_{12}}\right)^2
-\frac{1}{2}\left(\frac{\sin\frac{\pi}{2} x_{23}}{\frac{\pi}{2} x_{23}}\right)^2-\frac{1}{2}\left(\frac{\sin\frac{\pi}{2} x_{31}}{\frac{\pi}{2} x_{31}}\right)^2\nonumber\\
&+\frac{1}{2}\frac{\sin\frac{\pi}{2} x_{12}}{\frac{\pi}{2} x_{12}}\frac{\sin\frac{\pi}{2} x_{23}}{\frac{\pi}{2} x_{23}}\frac{\sin\frac{\pi}{2} x_{31}}{\frac{\pi}{2} x_{31}},
\end{align}
again in a weak sense.
\begin{figure}
\centering
\includegraphics[width=.4\columnwidth]{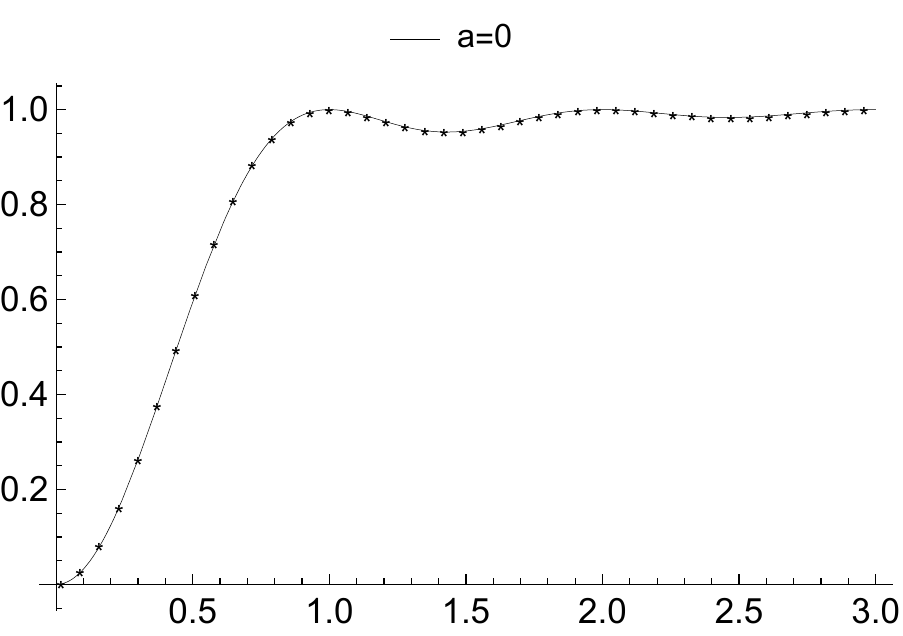}\qquad
\includegraphics[width=.4\columnwidth]{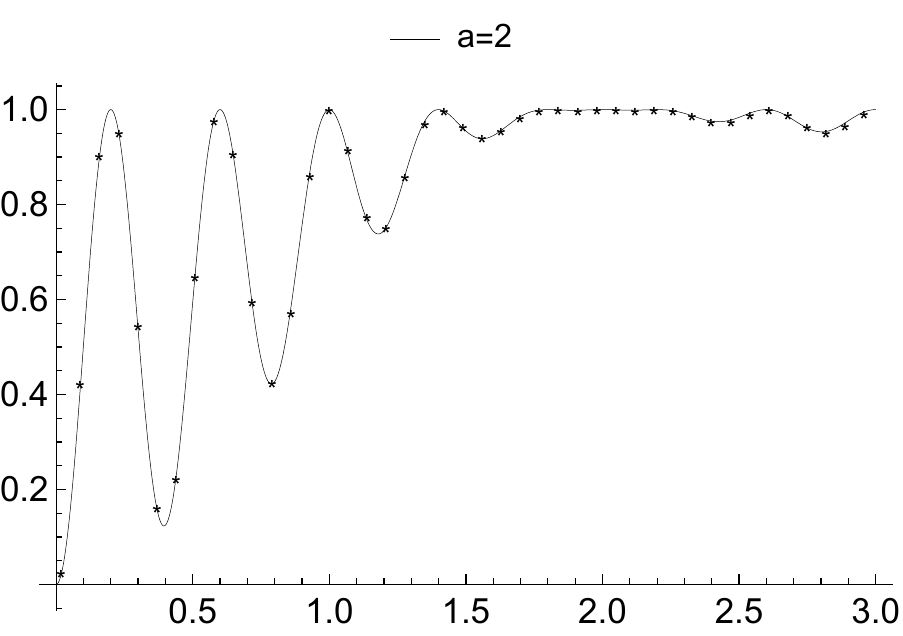}\\
\includegraphics[width=.4\columnwidth]{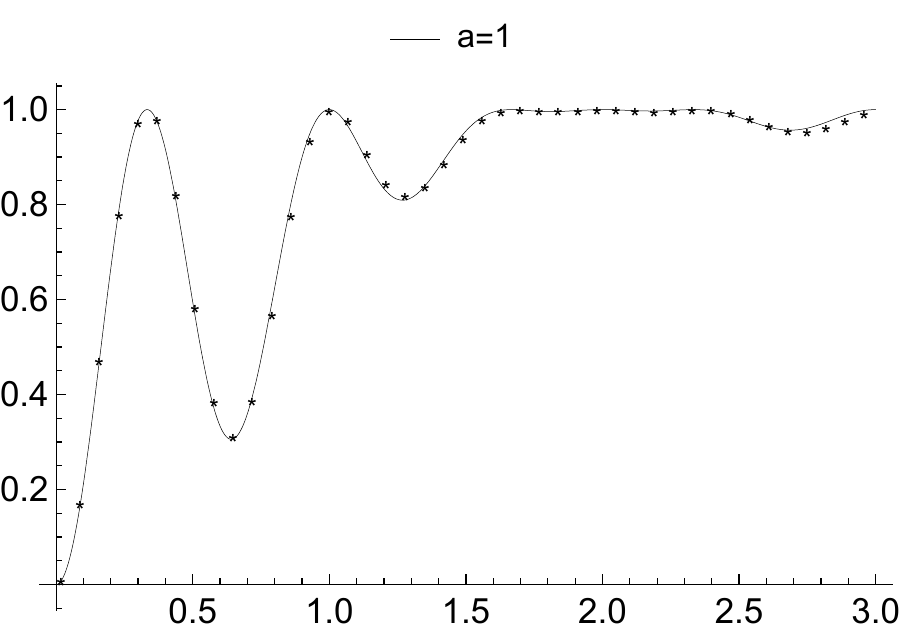}\qquad
\includegraphics[width=.4\columnwidth]{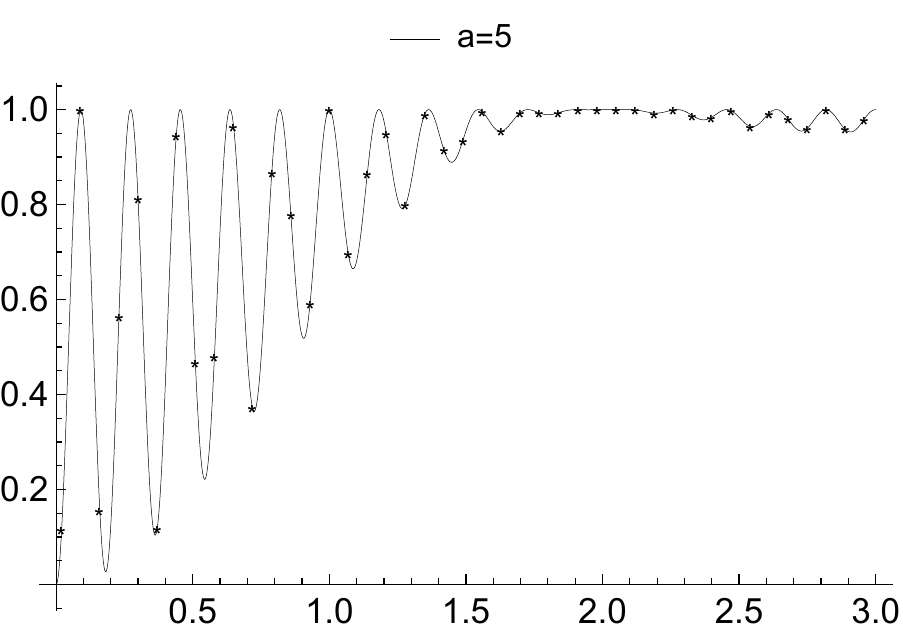}
\caption{Scaling limit of two-point correlation functions $\rho_1(0)^{-1}\rho_2(\rho_1(0)^{-1}x,\rho_1(0)^{-1}y)$,  as a function
of $|x-y|$,
for several values of $a$ (dots). Here $M=20$. The solid lines are the limits given in
Eq. \eqref{two_point.1}.}
\label{fig:2pas}
\end{figure}
This pattern can be generalised for  generic $n$ as follows. The $n$-point correlation function is given by the determinantal formula
\begin{align}
\tilde{\rho}_n(x_1,\dots,x_n)&=\det_{1\leq i,j\leq n}k(x_i-x_j)=\sum_{\sigma\in S_n}(-1)^{n-m(\sigma)}\prod_{i=1}^nk(x_{\sigma(i)}-x_i)\nonumber\\
&=\sum_{\sigma\in S_n}(-1)^{n-m(\sigma)}\prod_{i=1}^n\frac{\sin\frac{\pi}{2} (x_i-x_{\sigma(i)})}{\frac{\pi}{2} (x_i-x_{\sigma(i)})}\prod_{i=1}^n\cos \omega(x_{\sigma(i)}-x_i)
\end{align}
($m(\sigma)$ denotes the number of cycles in $\sigma\in S_n$).
For large $a$, the product of cosines becomes
\be
\lim_{\omega\to\infty}\prod_{i=1}^n\cos \omega(x_{\sigma(i)}-x_i)=\left(\frac{1}{2}\right)^{n-m(\sigma)}.
\ee
We remind the reader that this limit is in the weak sense of integration over compact subsets
of $\R^n$ or equivalently, integration against bounded measurable functions on $\R^n$
with compact support.
For the proof, one can use the addition formulae of the trigonometric functions or, alternatively, observe that, as $\omega\to\infty$,  
\begin{align}
\int_{0}^1\cos\omega(x-x)\de x&=1\\
\int_{0}^1 \int_{0}^1\cos^2\omega(x-y)\de x\de y&=\frac{1}{2}\left(1+\frac{\sin^2\omega}{\omega^2}\right)=\frac{1}{2}+o(1),\\
\int_0^1\cos\omega(x-z)\cos\omega(z-y)\de z&=\frac{1}{2}\cos\omega(x-y)+o(1).
\end{align}
Hence, for a given permutation $\sigma\in S_n$, we see that a cycle of length $m_i$ contributes to the product with a factor $(1/2)^{m_i-1}$. For example, each fixed point gives a factor $1$, each transposition gives a factor $1/2$, a $3$-cycle gives $1/4$, and so on. If the $m(\sigma)$ cycles of $\sigma$ have lengths $m_i$, $i=1,\dots,m(\sigma)$, 
\begin{align}
\prod_{i=1}^n\cos \omega(x_{\sigma(i)}-x_i)&=
\left(\frac{1}{2}+o(1)\right)^{m_1-1}\left(\frac{1}{2}+o(1)\right)^{m_2-1}\cdots\left(\frac{1}{2}+o(1)\right)^{m_{m(\sigma)}-1}\nonumber\\
&=\left(\frac{1}{2}\right)^{\sum_{i=1}^{m(\sigma)}(m_i-1)}+o(1)=\left(\frac{1}{2}\right)^{n-m(\sigma)}+o(1).
\end{align}
Therefore, as $a\to\infty$,
\begin{align}
\tilde{\rho}_n(x_1,\dots,x_n)&=\sum_{\sigma\in S_n}(-1)^{n-m(\sigma)}\prod_{i=1}^n\frac{\sin\frac{\pi}{2} (x_i-x_{\sigma(i)})}{\frac{\pi}{2} (x_i-x_{\sigma(i)})}\prod_{i=1}^n\cos\omega(x_{\sigma(i)}-x_i)\nonumber\\
&\stackrel{a\to\infty}{\to}\sum_{\sigma\in S_n}\left(-\frac{1}{2}\right)^{n-m(\sigma)}\prod_{i=1}^n\frac{\sin\frac{\pi}{2} (x_i-x_{\sigma(i)})}{\frac{\pi}{2} (x_i-x_{\sigma(i)})}.
\end{align}
in the sense that
\begin{multline}
\lim_{a\to\infty}\int\det_{1\leq i,j\leq n}k(x_i-x_j)f(x_1,\dots,x_n)\de x_1\cdots \de x_n=\\
\int\detalpha{1\leq i,j\leq n}{-1/2}\frac{\sin\frac{\pi}{2}(x_i-x_j)}{\frac{\pi}{2}(x_i-x_j)}f(x_1,\dots,x_n)\de x_1\cdots \de x_n,
\end{multline}
for any  $f(x_1,\dots,x_n)$ bounded, measurable function with compact support. This implies convergence of gap probabilities and number density (integrated over compact sets) hence, by Kallenberg's criteria~\cite[Theorem 4.5]{Kallenberg74}\cite[Theorem 3.3]{Kallenberg86},
 weak convergence of the associated point processes.
In particular, as $a\to\infty$, the process converges weakly to the union of two independent rescaled sine processes with kernel $\sin\frac{\pi}{2}(x-y)/(\frac{\pi}{2}(x-y))$. Note that this is not the standard sine kernel in the bulk of the ground state ($a=0$) which is $\sin(\pi(x-y))/(\pi(x-y))$.
\par
\subsection{Local statistics at the cusps and the edges} It is clear that the previous analysis holds for any fixed point $x_0\in\R$, where we have
\begin{align}
\lim_{M\to\infty}\rho_1(x_0)^{-1}K_J\left(x_0+\rho_1(x_0)^{-1}x,x_0+\rho_1(x_0)^{-1}y\right)
=\frac{\sin\frac{\pi}{2}(x-y)}{\frac{\pi}{2}(x-y)}\cos\omega(x-y),
\label{eq:transl}
\end{align}
with $\omega=\frac{\pi}{2}(2a+1)$.
A look at Fig.~\ref{fig:one-point} suggests that, for large $N$, the local correlations of the block projection process depend on the `region' where we take the scaling limit. There are two points in the support of the density that look special: the cusps at $\pm2a\sqrt{M}$ and the edges $\pm2(a+1)\sqrt{M}$. For example, it is clear that the local statistics at the edges cannot be described by a translation invariant kernel of the type~\eqref{eq:transl}. We can examine the scaling limits at points $x_0=2b\sqrt{M}$ not in the bulk. We report here the results (they follow from the know asymptotics~\eqref{eq:sinekernel}-\eqref{eq:Airykernel} of the GUE process and the block structure of $K_J(x,y)$):
\begin{itemize}
\item[(i)] (Before the cusp) Set $x_0=2b\sqrt{M}$ with $0\leq b\leq a$:
\begin{align}
\lim_{M\to\infty}\rho_1(x_0)^{-1}K_J\left(x_0+\rho_1(x_0)^{-1}x,x_0+\rho_1(x_0)^{-1}y\right)
=\frac{\sin\frac{\pi}{2}(x-y)}{\frac{\pi}{2}(x-y)}\cos\omega(x-y),
\label{eq:before_cusp}
\end{align}
with 
\be
\omega=\frac{\pi}{2}\frac{\sqrt{(a+1)^2-b^2}+\sqrt{a^2-b^2}}{\sqrt{(a+1)^2-b^2}-\sqrt{a^2-b^2}}.
\ee
At the cusp, i.e $b=a$, this is the sine kernel;
\item[(ii)] (After the cusp, before the edge) At $x_0=2b\sqrt{M}$ with $a\leq b< (a+1)$:
\begin{align}
\lim_{M\to\infty}\rho_1(x_0)^{-1}K_J\left(x_0+\rho_1(x_0)^{-1}x,x_0+\rho_1(x_0)^{-1}y\right)
=\frac{\sin\pi(x-y)}{\pi(x-y)};
\label{eq:after_cusp}
\end{align}
\item[(iii)] (At the edge) At $x_0=2(a+1)\sqrt{M}$, we take the `edge scaling' $N^{1/6}$:
\begin{align}
\lim_{M\to\infty}\frac{1}{N^{1/6}}K_J\left(x_0+\frac{x}{N^{1/6}},x_0+\frac{y}{N^{1/6}}\right)=\frac{\Ai(\eta x)\Ai'(\eta y)-\Ai'(\eta x)\Ai(\eta y)}{x-y},
\label{eq:at_edge}
\end{align}
with $\eta=\frac{(a+1)^{1/3}}{(2a+1)^{1/6}}$. This is just a rescaling of the Airy kernel.
\end{itemize}

We remark that, when $a>0$, between the cusps (i) the limit kernel depends explicitly on the bulk point $2b\sqrt{M}$ (as evident from~\eqref{eq:before_cusp}). This is very different from the `quantum bulk' (ii) where the scaling limit the kernel is always the sine kernel~\eqref{eq:after_cusp} (as long as we are not at the edges). In this sense, for $a>0$ there is a `classical bulk' regime which is absent in the ground state $a=0$. When $b\to a$ from below, the limit kernel freezes to the sine kernel and no longer depends on $b$.
It is worth noticing that this transition from~\eqref{eq:before_cusp} to~\eqref{eq:after_cusp} across the cusp $x_0=2a\sqrt{M}$ is continuous. 
We can also discuss the question of the matching in the limit of large $a$. 
When $a\to \infty$, if $b^2 = a^2 - 2\tau a$ with $\tau\geq0$, then $\omega\to c \pi/2$ ,
where $c = (1+\sqrt{\tau/(1+\tau)})/((1-\sqrt{\tau/(1+\tau)})$.  If $\tau=0$ (i.e. $b=a$) this gives the sine kernel;
$\tau\to\infty$ gives $\omega\to\infty$ and we have the $-1/2$-determinantal process discussed in the previous section.

So there is a family of kernels in between with a fixed $\pi/2 < \omega < \infty$, which are
seen just inside the cusp when $a\to\infty$; they are the 
same as if one is looking inside the bulk, between the cusps, and keeping $a$ fixed.
\section{Block projection processes}
\label{sec:block}
Let us summarise the limit theorems of the previous two sections in a slightly generalised setting. 
Consider the determinantal process with kernel (block projection)
\ben
K_J(x,y)=\sum_{k\in J}\overline{\psi_k}(x)\psi_k(y).
\een
Suppose that the set of energy levels is $J=\inter{a^2M}{(a+r)^2M}$, with $r$ positive integer.  There are two cases for the rescaled processes in the bulk:
\begin{itemize}
\item If $a=0$, then
\be
\lim_{M\to\infty} \rho_1(0)^{-n}\det_{1\leq i,j\leq n} K_J(\rho_1(0)^{-1}x_i,\rho_1(0)^{-1}x_j)=\detalpha{1\leq i,j\leq n}{-1}\frac{\sin\pi(x_i-x_j)}{\pi(x_i-x_j)};
\ee
\item If $a>0$, then
\be
\lim_{a\to\infty}\lim_{M\to\infty} \rho_1(0)^{-n}\det_{1\leq i,j\leq n} K_J(\rho_1(0)^{-1}x_i,\rho_1(0)^{-1}x_j)=\detalpha{1\leq i,j\leq n}{-\frac{1}{2}}\frac{\sin\frac{\pi}{2}(x_i-x_j)}{\frac{\pi}{2}(x_i-x_j)}
\ee
\end{itemize}
\par
In this Section we set to ourselves to find a suitable limit procedure to obtain $\alpha$-determinantal processes out of $K_J(x,y)$ with $\alpha=-\frac{1}{m}$, with $m$ generic positive integer.

From the previous analysis we understand that a key ingredient to obtain non-trivial scaling limits is the possibility to rearrange $K_J(x,y)$ as a sum of Christoffel-Darboux kernels. Let us consider a subset $J$ of energy levels with a block structure (the union of $B$ blocks)
\begin{align}
J&=\bigcup_{j=0}^{B-1}\inter{a_j^2M}{(a_j+r_j)^2M}
\end{align}
Hereafter, we assume that the $a_j$'s and $r_j$'s are such that  $J$ is a union of $B$ \emph{disjoint} blocks. The number of energy levels $N=|J|$ is
\be
N=\sum_{j=0}^{B-1}(a_j+r_j)^2-a_j^2.
\ee
Denote by $\Psi_J$ the wave function representing  $N$ fermions with one fermion in each level $k\in J$. In formulae,
\be
\Psi_J(x_1,\cdots,x_N)=\frac{1}{\sqrt{N!}}\det_{1\leq i,j\leq N} \psi_{k_i}(x_j),\qquad \text{with $k_i\in J$}.
\label{eq:wf_J}
\ee
Then,
\be
\left|\Psi_J(x_1,\cdots,x_{N})\right|^2=\frac{1}{N!}\det_{1\leq i,j\leq  N} K_J(x_i,x_j)
\ee
defines a determinantal point process on the line with kernel $K_J(x,y)$.
\par
When $M$ is large (a limit of large number of particles) the one-point function is
\be
\rho_1(x)\stackrel{M\to\infty}{\sim} \frac{1}{2\pi}\sum_{j=0}^{B-1}\left(\sqrt{(4(a_j+r_j)^2 M-x^2)_+}-\sqrt{(4a_j^2M-x^2)_+}\right).
\label{eq:semicirc_annuli}
\ee
In the bulk, e.g. at $x=0$, 
\be
\lim_{M\to\infty}\frac{1}{\sqrt{M}}\rho_1(0)=\frac{R}{\pi},\quad \text{with\,\,$
R=\sum_{j=0}^{B-1}r_j$}.
\ee
It is not difficult to verify that the previous semiclassical considerations for the one-point function based on the correspondence principle (see Eq.~\eqref{eq:Wigner_f}) carry over in the case of several blocks. For $B>1$, when $M\to\infty$ the Wigner function in the phase space is uniform on $B$ nested annuli; the projection onto the real line of the uniform density on nested annuli gives the number density~\eqref{eq:semicirc_annuli}. See Fig.~\ref{fig:diffr}.
\par
The scaling limit of the kernel in the bulk is
\be
\lim_{M\to\infty}\frac{1}{\rho_1(0)}K_J\left(\frac{x}{\rho_1(0)},\frac{y}{\rho_1(0)}\right)=k(x-y),
\ee
with
\begin{align}
k(x-y)=\sum_{j=0}^{B-1}\frac{\sin\left(\frac{\pi r_j(x-y)}{2R}\right)}{\frac{\pi r_j(x-y)}{2R}}\frac{r_j}{R}\cos\frac{\pi (2a_j+r_j)(x-y)}{2R}.
\end{align}
There are two special block structures that give rise to $\alpha$-determinantal processes, $\alpha=-\frac{1}{m}$ with  $m$ even or odd.

\subsection{$J$ of even type and $-\frac{1}{2B}$-determinantal processes}
\begin{figure}
\centering
\includegraphics[width=1\columnwidth]{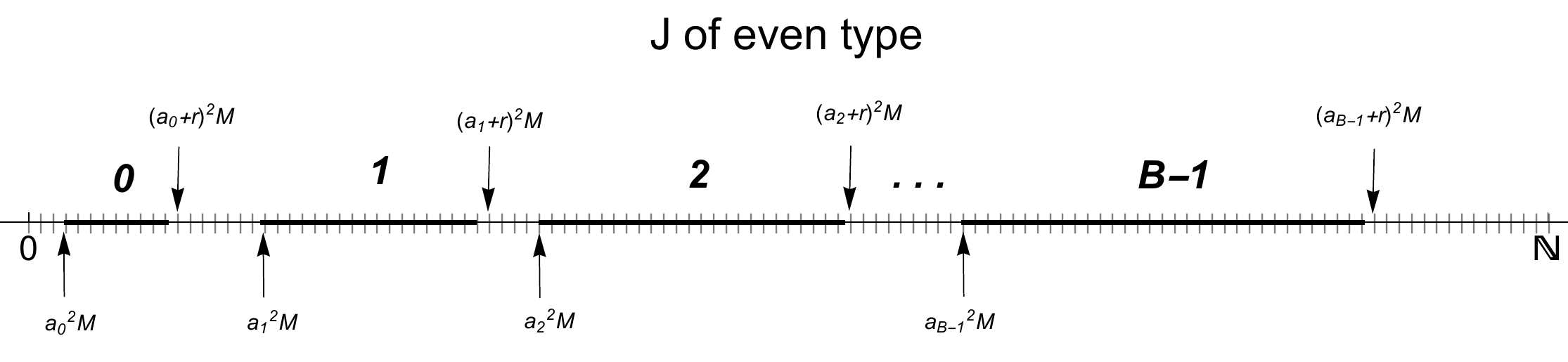}\\\vspace{.5mm}
\includegraphics[width=1\columnwidth]{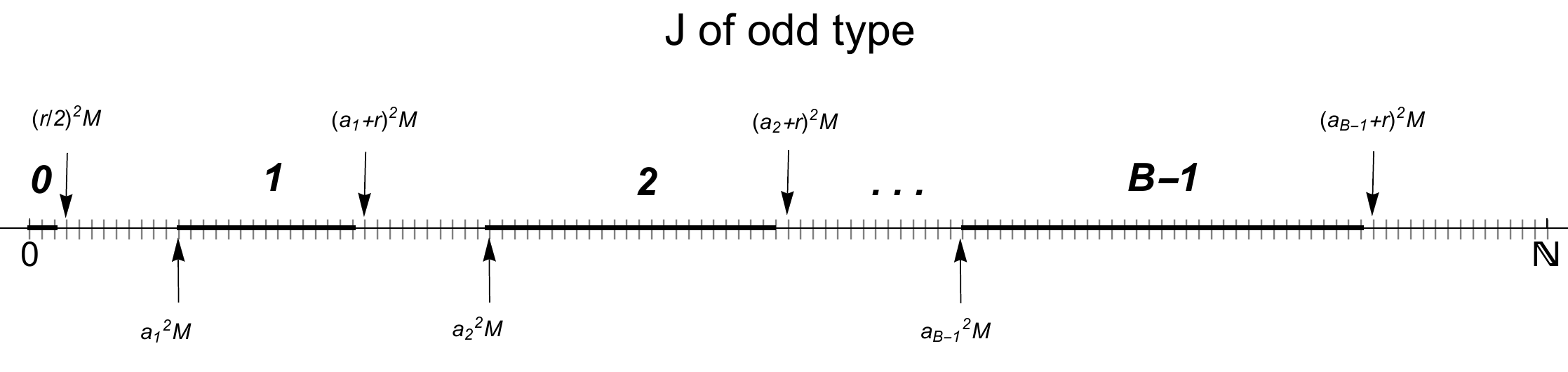}
\caption{Scheme of the even type (top) and odd type (bottom) of subsets $J$'s .}
\label{fig:scheme}
\end{figure}
Suppose that $0<a_0<a_1<\cdots<a_{B-1}$, and choose $r_0=r_1=\cdots=r_{B-1}=r$, so that $R=rB$. See top panel of Fig.~\ref{fig:scheme}. In formulae.
\be
J=\inter{a_0^2M}{(a_0+r)^2M}\cup\inter{a_1^2M}{(a_1+r)^2M}\cup\cdots\cup\inter{a_{B-1}^2M}{(a_{B-1}+r)^2M}
\ee 
We say, for shortness, that $J$ is of \emph{even} type.
Then, 
\begin{align}
k(x-y)=\frac{\sin\left(\frac{\pi (x-y)}{2B}\right)}{\frac{\pi (x-y)}{2B}}\cdot\frac{1}{B}\sum_{j=0}^{B-1}\cos\omega_j(x-y),\quad\text{where $\omega_j=\frac{\pi\left(2a_j+r\right)}{2rB}$}.
\end{align}

If $a_0,\dots,a_{B-1}$ are sent (independently) to infinity, then for any $\sigma\in S_n$,
\be
\lim_{a_0,\dots,a_{B-1}\to\infty}\prod_{i=1}^n\frac{1}{B}\sum_{j=0}^{B-1}\cos \omega_j(x_{\sigma(i)}-x_i)=\left(\frac{1}{2B}\right)^{n-m(\sigma)}.
\ee
To see this, we expand the sum to get
\begin{align}
\prod_{i=1}^n\frac{1}{B}\sum_{j=0}^{B-1}\cos \omega_j(x_{\sigma(i)}-x_i)&=
\frac{1}{B^n}\sum_{j_1,\dots,j_n=0}^{B-1}\cos \omega_{j_1}(x_{\sigma(1)}-x_1)\cdots \cos \omega_{j_n}(x_{\sigma(n)}-x_n)\nonumber \\
&=
\frac{1}{B^n}B^{m(\sigma)}\left(\frac{1}{2}\right)^{m_1-1}\cdots\left(\frac{1}{2}\right)^{m_{m(\sigma)}-1}+o(1),
\end{align}
where the factor $B^{m(\sigma)}$ is the number of ways of assigning frequencies $\omega_j$, $j=0,\dots,B-1$ to the  $m(\sigma)$ cycles of $\sigma$.
We conclude that, if $J$ is of even type, then
\be
\lim_{\substack{a_0,\ldots,a_{B-1}\to\infty}}\lim_{M\to\infty} \det_{1\leq i,j\leq n}  \frac{1}{\rho_1(0)^{n}}\det_{1\leq i,j\leq n} K_J\left(\frac{x_i}{\rho_1(0)},\frac{x_j}{\rho_1(0)}\right)=\detalpha{1\leq i,j\leq n}{-\frac{1}{2B}}\frac{\sin\frac{\pi}{2B}(x_i-x_j)}{\frac{\pi}{2B}(x_i-x_j)}.
\label{eq:2B}
\ee
\par
\subsection{$J$ of odd type and $-\frac{1}{2B-1}$-determinantal processes}
Suppose now that $0=a_0<a_1<\cdots<a_{B-1}$, and  choose $2r_0=r_1=\cdots=r_{B-1}=r$, i.e., 
\be
J=\inter{0}{(r/2)^2M}\cup\inter{a_1^2M}{(a_1+r)^2M}\cup\cdots\cup\inter{a_{B-1}^2M}{(a_{B-1}+r)^2M}
\ee 
so that $R=r(B-1/2)$. We say that $J$ is of \emph{odd} type. See bottom panel of Fig.~\ref{fig:scheme}. Then,
\begin{align}
k(x-y)=\frac{\sin\left(\frac{\pi (x-y)}{2B-1}\right)}{\frac{\pi (x-y)}{2B-1}}\cdot\frac{1}{B-\frac{1}{2}}\left(\frac{1}{2}+\sum_{j=1}^{B-1}\cos\omega_j(x-y)\right),\quad\text{where $\omega_j=\frac{\pi\left(2a_j+r\right)}{2r(B-\frac{1}{2})}$}.
\end{align}
\par
One can check that, in the limit of large $a_1,\dots,a_{B-1}$, for any $\sigma\in S_n$,
\be
\lim_{a_1,\dots,a_{B-1}\to\infty}\prod_{i=1}^n\frac{1}{B-\frac{1}{2}}\left(\frac{1}{2}+\sum_{j=1}^{B-1}\cos \left(\omega_j(x_{\sigma(i)}-x_i)\right)\right)=\left(\frac{1}{2B-1}\right)^{n-m(\sigma)}.
\ee
The conclusion is that, if $J$ has $B$ blocks and is of odd type, then
\be
\lim_{\substack{a_1\ldots,a_{B-1}\to\infty}}\lim_{M\to\infty} \det_{1\leq i,j\leq n}  \frac{1}{\rho_1(0)^{n}}\det_{1\leq i,j\leq n} K_J\left(\frac{x_i}{\rho_1(0)},\frac{x_j}{\rho_1(0)}\right)=\detalpha{1\leq i,j\leq n}{-\frac{1}{2B-1}}\frac{\sin\frac{\pi}{2B-1}(x_i-x_j)}{\frac{\pi}{2B-1}(x_i-x_j)}.
\label{eq:2B-1}
\ee

\section{Summary and remarks}
\label{sec:sum}
We can summarise the findings of the previous sections as follows.
\begin{thm} 
Let $K_J(x,y)$ be a kernel where $J$ has $B$ blocks as above. Consider the block projection process with correlation kernel  ${\rho_1(0)^{-1}}K_J\left({\rho_1(0)^{-1}}x,{\rho_1(0)^{-1}}y\right)$. Then, in the limit $M\to\infty$ (first) and $a_i\to\infty$, the process converges to the $\alpha$-determinantal process with kernel $\frac{\sin\pi\alpha(x-y)}{\pi\alpha(x-y)}$ (the union of $-1/\alpha$ rescaled sine processes). The parameter $\alpha<0$ is $\alpha=-\frac{1}{2B}$ or $\alpha=-\frac{1}{2B-1}$ depending on whether $J$ is of even or odd type, respectively. The convergence is in the sense of weak convergence of point processes.
\end{thm}
\par
Note that the correlation functions $\rho_n(x_1,\dots,x_n)=\detalpha{1\leq i,j\leq n}{\alpha}\frac{\sin\pi\alpha(x-y)}{\pi\alpha(x-y)}$ are bounded
\be
\rho_n(x_1,\dots,x_n)\leq 1,
\ee
and hence determine uniquely the point process~\cite{Lenard73}. This limit process is translation invariant, and standard quantities of interest in the theory of point processes can be investigated. 
\par
\subsection{Pair statistics and number variance} The pair statistics in Fourier space is traditionally studied by looking at properties of the \emph{structure factor} defined (for a process with unit density) as~\cite{Scardicchio09,Torquato08}
\ben
S(k)=1+\hat{h}(k),
\een
where $\hat{h}(k)$ is the Fourier transform of the \emph{total} or \emph{connected correlation function}
\ben
h(r)=\rho_2(x_1,x_2)-1,\qquad r=x_1-x_2.
\een
For the process with correlation functions $\detalpha{1\leq i,j\leq n}{\alpha}\frac{\sin\pi\alpha(x-y)}{\pi\alpha(x-y)}$ it is easy to calculate
\be
S(k)=
\begin{cases}
\frac{|k|}{2\pi|\alpha|}&\text{if $|k|\leq 2\pi|\alpha|$}\\
1&\text{if $|k|> 2\pi|\alpha|$}
\end{cases}.
\ee
\par
We remark that, as the number of blocks $B$ increases, we obtain a Poisson process, as expected (superposition of a large number of independent spectra~\cite{Berry84}). Indeed, when 
$B\to\infty$, $\alpha\to 0$ and $\frac{\sin\pi\alpha(x-y)}{\pi\alpha(x-y)}\to 1$ for all $x$ and $y$. Consequently,
$\rho_2(x,y)\to 1$ and hence $h(r)\to 0$, leading to $S(k)=1$ (the structure factor of a Poisson process).
\par
\begin{figure}
\centering
\includegraphics[width=.7\columnwidth]{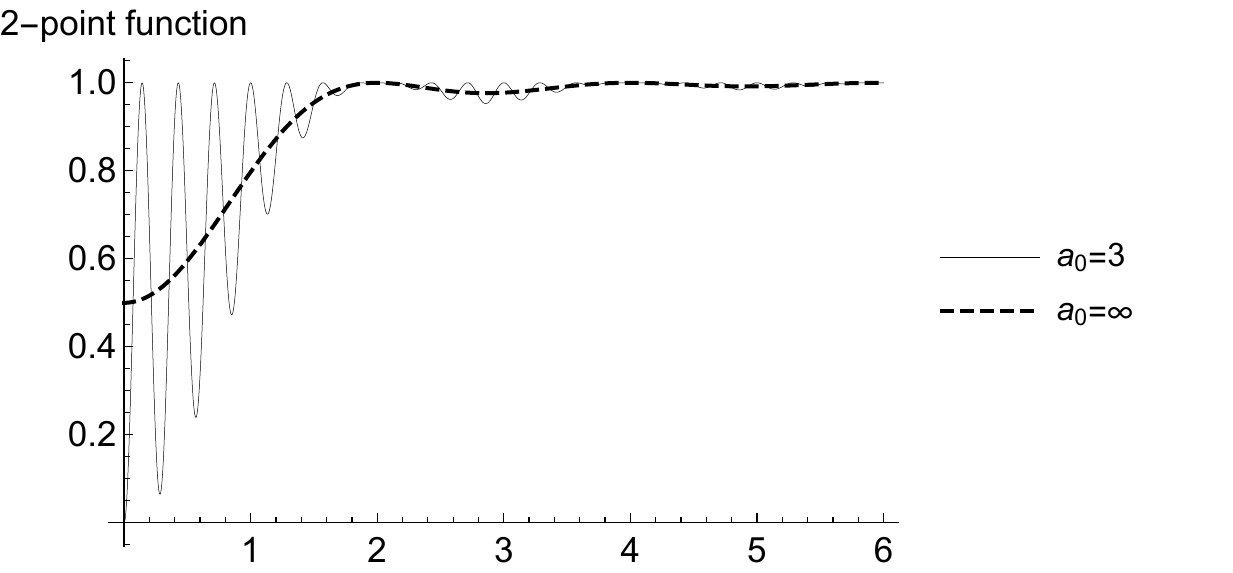}
\caption{Comparison of the two-point correlation function of a block projection process and its limit. Here $B=1$, and the limit process is $\alpha$-determinantal with $\alpha=-1/2$.}
\label{fig:frequency}
\end{figure}
\begin{figure}
\centering
\includegraphics[width=.7\columnwidth]{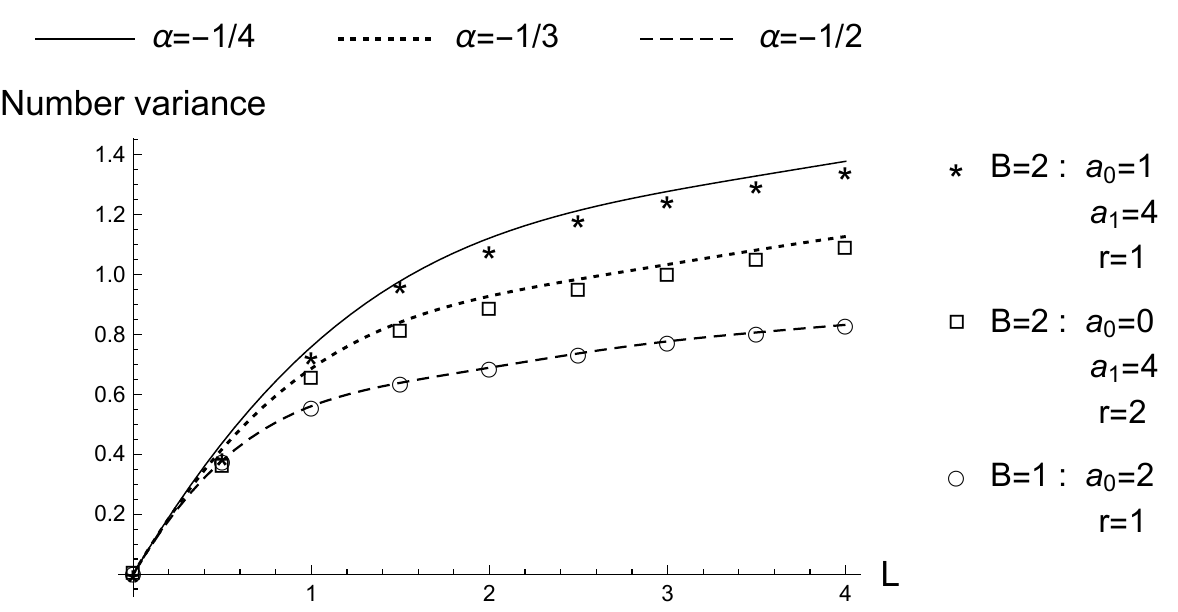}
\caption{Number variance. The symbols comes from numerical integrations of~\eqref{eq:var_num_k} while the lines are the corresponding limits~\eqref{eq:var_num_alpha}. Even for moderate values of the $a$'s the agreement is fairly good.}
\label{fig:num_var}
\end{figure}

As already discussed, the convergence of the correlation functions when $a_i\to\infty$ is not pointwise. This is quite clear, as the cosine factors in $k(x-y)$ oscillates with high frequency. See Fig.~\ref{fig:frequency}.  
To illustrate better this point we consider the \emph{number variance}, i.e. the variance of the number of fermions in a box $[-L/2,L/2]$ in the bulk when the quantum state of the fermions is $\Psi_J(x_1,\dots,x_N)$. The expected number of particles is
\be
E\left(\#\left[-\frac{L}{2},\frac{L}{2}\right]\right)=\int_{-L/2}^{L/2}K_{J}(x,x)\de x.
\ee
In the scaling limit in the bulk, the process becomes translation invariant and
\be
\lim_{M\to\infty}E\left(\#\left[-\frac{L}{2\rho_1(0)},\frac{L}{2\rho_1(0)}\right]\right)=L.
\ee
Standard manipulations give a formula for the variance in terms of the kernel $K_J(x,y)$:
\be
\operatorname{Var}\left(\#\left[-\frac{L}{2},\frac{L}{2}\right]\right)=\int_{-L/2}^{L/2}K_{J}(x,x)\de x-\iint_{-L/2}^{L/2}K_{J}(x,y)^2\de x\de y.
\ee
Taking the limit $M\to\infty$,
\be
\lim_{M\to\infty}\operatorname{Var}\left(\#\left[-\frac{L}{2\rho_1(0)},\frac{L}{2\rho_1(0)}\right]\right)=L-\iint_{-L/2}^{L/2}k(x-y)^2\de x\de y,
\label{eq:var_num_k}
\ee
and, for large $a_i$'s, the weak convergence of the process implies
\begin{align}
\lim_{a_i\to\infty}\lim_{M\to\infty}\operatorname{Var}\left(\#\left[-\frac{L}{2\rho_1(0)},\frac{L}{2\rho_1(0)}\right]\right)&=L+\alpha\iint_{-L/2}^{L/2}\left(\frac{\sin\alpha\pi(x-y)}{\alpha\pi(x-y)}\right)^2\de x\de y
\label{eq:var_num_alpha}
\end{align}
where $\alpha=-\frac{1}{2B}$  or  $\alpha=-\frac{1}{2B-1}$, if $J$ is of even or odd type, respectively. 
For an illustration of this convergence, see Fig.~\ref{fig:num_var}.
\par
In particular, in the limit $M\to\infty$ and $a_i\to\infty$, the number variance has the asymptotic expansions
\be
\operatorname{Var}\left(\#\left[-\frac{L}{2\rho_1(0)},\frac{L}{2\rho_1(0)}\right]\right)\sim
\begin{cases}
L+\alpha  L^2-\frac{1}{18} \pi ^2 \alpha ^3 L^4+\frac{2  }{675} \pi ^4 \alpha ^5L^6+\cdots&\text{as $L\to0$}\\\\
-\dfrac{1}{\alpha\pi^2}\left(\log L+\log(-2\pi\alpha)+1+\gamma_E+\cdots\right)&\text{as $L\to\infty$}
\end{cases},
\ee
where $\gamma_E=0.577215\dots$ is the Euler-Mascheroni constant. For $\alpha=-1$, the second line
reduces to the well-known Dyson-Mehta result for the GUE~\cite{Mehtabook}.

\subsection{Heuristic discussion and extension to other models} At this stage one may ask for a semiclassical explanation of the convergence of the fermion processes to $\alpha$-determinantal processes. 
It is known that fermions generically display `Friedel oscillations'~\cite{Friedel52,Friedel58,Gleisberg00} in the particle density and correlation functions with a wave vector determined by a combination of Fermi surface effects and many-body effects. In the one-dimensional setting of non-interacting particles considered in this paper, the oscillations described by the sine kernel are simply a consequence of the sharpness of the Fermi surface (here points) at zero temperature. In the ground state, the momenta are in the Fermi sphere (interval) with edges $\pm\sqrt{N}$, and in the bulk the correlation kernel is the Fourier transform of the indicator function of that interval, hence the sine kernel with frequency $(1/2)\sqrt{N}$.

\begin{figure}
\centering
\includegraphics[width=.8\columnwidth]{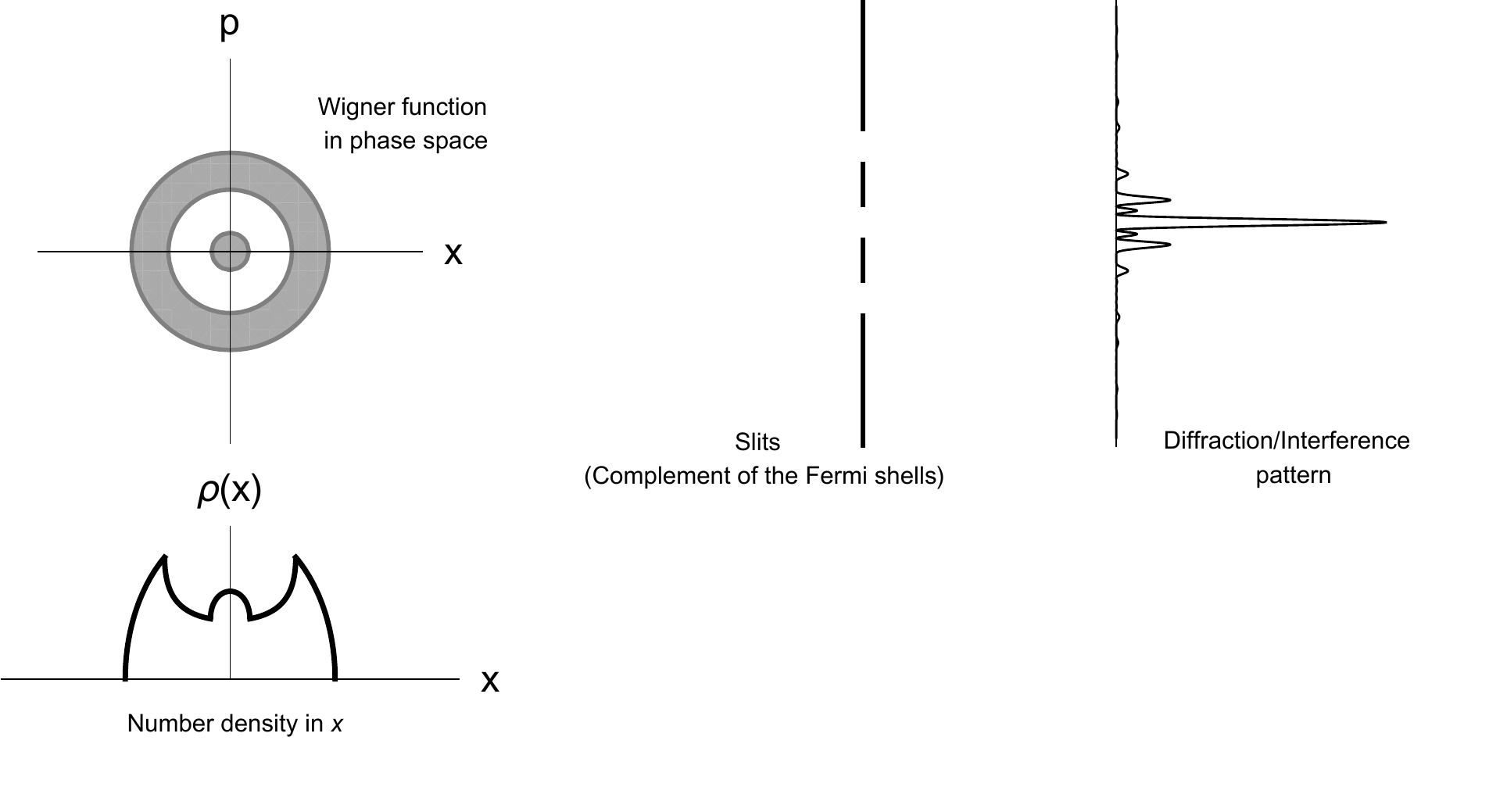}
\caption{Scheme of the semiclassical considerations based on the correspondence principle. For large $M$, the Wigner function is uniform on a set of nested annuli. The projection onto the $x$-axis is the limit number density. The intersections of the annuli with the $p$-axis define the Fermi shells. The Fourier transform of the Fermi shells gives a sum of sine processes that become independent when the shells are far apart. A close analogy can be drawn with the diffraction patterns in wave optics.}
\label{fig:diffr}
\end{figure}

For $B>1$ blocks, the Fermi sphere, i.e. the set of momenta $p\in\R$ for the wavefunction $\Psi_J$, is rather the union of Fermi shells (disjoint intervals). Oscillations occur in the correlation functions in the bulk, and their frequencies is related to the size of the Fermi shells.  More precisely, if $J$ is of even type, the set of possible momenta consists of $2B$ shells  symmetric with respect to $p=0$. If $J$ is of odd type, there are $2B-1$ intervals (one containing the origin) of possible values for the momenta. In both cases, when $M\to\infty$ each interval has the same length $\sim r\sqrt{M}$. This also explains why in the odd type we choose $r_0=r/2$. In the scaling limit in the bulk, to each Fermi shell corresponds a correlation kernel with frequency given by $(1/2)\times r\sqrt{M}$; for large $a_i$'s, the distance between the Fermi shells increases, and the oscillations of the kernels are asymptotically independent so that the process in the bulk becomes a superposition of independent sine processes with the same frequency. A glance at Fig.~\ref{fig:diffr} may be helpful. 

In fact, the reader may have recognised in the computation of the kernel $k(x-y)$ steps similar to the calculation of diffraction/interference patterns in wave optics~\cite{Sommerfeld}. For a single slit of width $r$ (ground state $a=0$) the far-field intensity distribution is proportional to $(\frac{\sin{\pi r z}}{\pi r z})^2$. For two slits of width $r$ at distance $a$ ($J$ of even type with one block) the interference pattern shows periodic fringes superimposed to the diffraction pattern $(\frac{\sin{\pi r z}}{\pi r z})^2\cos^2(\pi a)$
However, if the slits are too far apart (i.e. when $a\to\infty$), the waves coming from the two slits do not interfere, no fringes will be seen and the intensity distribution will be just the incoherent sum of the diffraction patterns from each individual slit. This easily extends to a generic number of slits.

These semiclassical considerations are also relevant in other block projection processes with correlation kernel
\ben
K_J(x,y)=\sum_{k\in J}\overline{\psi_k}(x)\psi_k(y),
\een
where $\psi_k$ form an orthonormal basis of some $L^2$ space. For instance, one can consider the free fermions on the circle, i.e. block projection processes constructed using the family of trigonometric polynomials $\psi_k(x)=e^{ikx}$, $k\in\mathbb{Z}$.
(These processes appeared under the name of `Fermi shell models'  in a work by Torquato, Scardicchio and Zachary~\cite{Torquato08}.) It is easy to see that the scaling limit in the bulk, in the limit of blocks very far apart the is a  superposition of independent sine processes.

As an alternative heuristics, one can imagine the Hermite block projection process as a complex Hermite block projection process conditioned to be real~\cite{Ledoux08}. This gives a nice heuristic explanation as to why we see a superposition of independent sine processes in the limit when we are inside the inner radius
of the annulus, coming from above and below (and becoming independent when $a\to\infty$). Similarly, in the circular case one should consider the block Ginibre process constructed using monomials $z^k$, $k\in\mathbb{Z}$;
then, constrained to the unit circle this is the block trigonometric process, and we see asymptotic superposition of sine processes as expected.

\subsection{Another $\alpha$-determinantal process from random matrices of finite size} 
The $\alpha$-determinantal processes described in this paper arise as scaling limits of block projection processes.
In particular, the limit processes describe configurations of an infinite number of particles (superposition of sine kernels). It is natural to ask whether it is possible to get (in a non-trivial way) $\alpha$-determinantal processes out of eigenvalues of random matrices of \emph{finite} size. In fact, one example of such a construction can be read off from an intriguing decoupling phenomenon for power of random unitary matrices discovered by Rains~\cite{Rains97,Rains03}. Let $m$ and $N$ 
be a positive integers with $m\leq N$, and let $U$ be a random unitary matrix from the Haar measure on $\mathcal{U}(mN)$. Then, the eigenvalues of $U^{m}$ are exactly distributed as the union of eigenvalues of $m$ independent unitary matrices $U_1,\dots,U_m$ chosen in $\mathcal{U}(N)$.
\par
It is a classical fact that the eigenvalues of random unitary matrices form a determinantal process on the unit circle. Set $S_n(z)=\frac{1}{2\pi}\frac{\sin(nz/2)}{\sin(z/2)}$, and denote by $x_1,\dots, x_{mN}$  the eigenphases of a random unitary $U$ of size $mN$. Then, the law of $x_1,\dots, x_{mN}$  defines a determinantal process with kernel $S_{mN}(x-y)$.
Rains' theorem can be restated by saying that the point configuration of $m$-th powers $x_1^m,\dots, x_{mN}^m$ is the union of $m$ independent determinantal processes with kernel $S_{N}(x-y)$.
Alternatively - and this is perhaps not so well-known -  the $m$-th powers $x_1^m,\dots, x_{mN}^m$ form an $\alpha$-determinantal process with $\alpha=-\frac{1}{m}$ and kernel $S_{mN}(x-y)$.
\par
Similar results hold for the eigenvalue processes of the other classical compact groups~\cite{Rains03}, and have been recently extended to a class of rotation invariant determinantal processes in the complex plane by Dubach~\cite{Dubach18}. It remains an open problem to generalise this construction to other matrix ensembles without rotation symmetry.

\ack
The research of FDC  and  NO'C is supported by ERC Advanced Grant 669306.
The research of FDC is partially supported by Gruppo Nazionale di Fisica Matematica (GNFM-INdAM).
\par

\end{document}